\documentclass[12pt]{iopart}
\usepackage{epsf} 
\begin{document}

\title[Percolation on the average and Potts model on graph]{Percolation on 
the average and spontaneous magnetization for q-states Potts model on graph}
\author{A. Vezzani 
\footnote{vezzani@fis.unipr.it}} 
 
\address{Dipartimento di Fisica, Istituto Nazionale di Fisica della Materia
(INFM), Universit\`a di Parma, parco Area delle Scienze 7A 43100 Parma Italy}

\begin{abstract}
We prove that the q-states Potts model on graph is 
spontaneously magnetized at finite temperature if and only if the 
graph presents percolation on the average.
Percolation on the average is a combinatorial problem defined by averaging
over all the sites of the graph the probability of belonging to a cluster of a
given size. In the paper we obtain an inequality between this average 
probability and the average magnetization, which is a typical extensive function
describing the thermodynamic behaviour of the model. 
\end{abstract}

 
\submitto{\JPA}


\section{Introduction}

The interplay between spin models and
percolation has been put into evidence since the fundamental work of Fortuin and
Kasteleyn \cite{F1}, where it is
shown that percolation is the limit for $q\to 1$ of the random 
cluster representation of q-states Potts model. This representation, as
already pointed out in \cite{F1}, is very general and can be performed 
on any discrete structure, i.e. on graph.

A further step has been the result proven in \cite{Aiz}, showing that, on 
lattices, the existence of percolation implies Potts transition for any value 
of $q$. The main purpose of this work is
the extension of the result to generic networks. 

An important issue in this direction has been a mathematical paper
\cite{Hagg} proving that Potts model has more than one Gibbs measure if
it is defined on graphs showing percolation. The
multiplicity of Gibbs measures is the most used definition of symmetry 
breaking in mathematical literature. However, it does not always
correspond to the thermodynamic concept of phase transition. 

Indeed, in the study of
the thermodynamic properties of physical systems, we are 
interested in the free energy and its derivatives 
such as the average magnetization, the susceptibility, the specific heat,
which are typical average extensive quantities. On the other hand, the 
existence of a single Gibbs measure, as explained in \cite{Hagg},
depends on a local parameter, that is the magnetization of a site of the graph.  
On lattices, translation invariance implies that local and average
quantities coincide (the proof of \cite{Aiz} depends on this fact). However, for
inhomogeneous structures the results can be different in the two cases. For
example, on the brush graph (see Figure 1) as we will prove in detail, 
there is not a single Gibbs measure, but the thermodynamic
functions do not present discontinuities and the spontaneous magnetization is 
zero.

\begin{figure}
\begin{center}
\leavevmode
\hbox{%
\epsfysize=6cm
\epsfbox{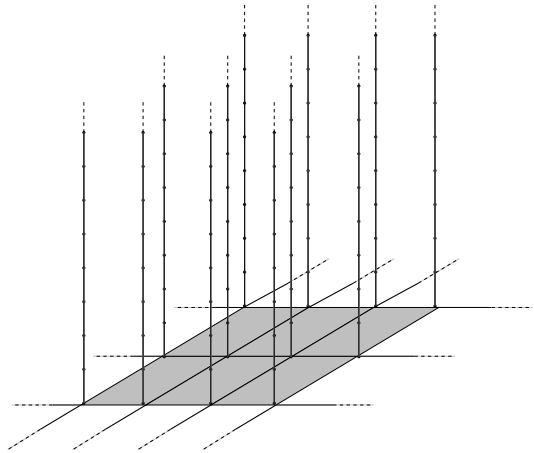}}
\caption{The brush graph is obtained by adding a linear chain 
to each site of a two-dimensional lattice.}
\end{center}
\end{figure}

A generalization of the result \cite{Aiz}, which takes into
account the thermodynamic behaviour of the system, requires the introduction 
of the new problem of percolation on the average. A graph $G$ is said to
present percolation on the average if the average, over all sites, of the 
probability of belonging to a
cluster of size greater than $l$ remains positive when $l\to\infty$. 
We will prove that the q-states Potts model on a graph $G$ is 
spontaneously magnetized  if and only if $G$ presents percolation on the 
average.

Therefore, the study of spin models with discrete symmetry leads to
a classification of graph topology which can be formulated 
in terms of a purely combinatorial problem. 
 
An analogous result \cite{MWFSSgen} has been previously obtained for classical 
spin models with continuous symmetry, 
where the thermodynamic behaviour depends on a different
combinatorial problem, i.e. random walks. In particular, the models present
spontaneous magnetization if and only if the graph is transient on the average
\cite{rimtim} (i.e. a graph where the average probability for a random-walk of
ever returning to the starting site is smaller than 1). 

This two classifications are different, indeed, it is well known
that lattices presenting percolation (i.e. lattices with dimension 
$d\geq 2$) are transient on the average only if $d\geq 3$. 
In recent works \cite{carpet}, 
it has been proved that the "anomalous" behaviour of presenting 
average percolation in a recurrent on the average structure is typical not only
of two dimensional lattices, but also of low dimensional networks, such as
the Sierpinski carpet. Therefore, a general result clarifying the relation 
between the two classifications is still lacking. Furthermore, up to now we do
not know if percolation on the average is related to graph topology 
by a simple parameter, as it is the spectral dimension \cite{specdim} in the 
case of transience on the average. 

The paper is organized as follows. In the next section we introduce the
Potts model and percolation on graph, we define percolation on the average and
we state the main result of the paper, i.e.  
Potts model is spontaneously magnetized if and only it is defined on a 
graph presenting percolation on the average. In the last 2 sections 
we prove the theorem.

\section{The q-states Potts model and percolation on graph}

A graph $G$ (see e.g. \cite{harari} for an introduction to graph theory) 
is a countable set $V$ of vertices or sites $i$ connected pairwise by a 
set $E$ of unoriented edges or links $(i,j)=(j,i)$. Connected sites 
are called nearest neighbours and we denote with $z_i$ the 
connectivity (number of neighbours) of the site $i$.
A path is a sequence of consecutive links $\{(i,k)
(k,h)\dots(n,m)(m,j)\}$. The chemical distance
$r_{i,j}$ is the length (number of links) of the shortest path
connecting the sites $i$ and $j$. 
The Van Hove sphere $S_{o,r}$ of center $o$ and radius 
$r$ is the set of the sites of $G$ such that $r_{o,i}\leq r$.  
We will call $E_{o,r}$ the set of all $(i,j)\in E$ such that $i,j \in S_{o,r}$,
$\partial S_{r}$ the set of the vertices of
$S_{r}$ connected with the rest of the graph
and we denote with $|\cdot|$ the cardinality of a set.
Let $\phi_i$ be a real function of the sites of an infinite graph,
the average in the thermodynamic limit $\bar{\phi}$ is 
\begin{equation}
\overline{\phi}\equiv 
\lim_{r\rightarrow\infty} 
{\displaystyle \sum_{i\in S_{o,r}} \phi_i \over \displaystyle |S_{o,r}|}~~.
\label{deftd}
\end{equation}
Interesting properties of the thermodynamic average, such as the independence
of the choice of the center $o$, are proved in \cite{rimtim}. 
Therefore, in the following we will drop the index $o$.
The measure $||S||$ of a subset $S$ of 
$V$ is the average value $\overline{\chi(S)}$ of its characteristic 
function $\chi_i(S)$ defined by
$\chi_i(S)=1$ if $i\in S$ and $\chi_i(S)=0$ if $i\not\in S$.

Important constraints on graph topology follow from the requirement to 
describe a physical network. Real systems, indeed, have bounded local energy 
and they are embedded in a finite dimensional space. Therefore, 
we will consider only connected graphs with bounded 
connectivity ($z_i<z_{max}$ $\forall i\in E$) and such that:
\begin{equation}
\lim_{r\to \infty} {|\partial S_{r}|\over |S_{r}|}~=~0
\label{isop} 
\end{equation}

The definition of the q-states Potts model 
(see e.g. \cite{potts} for a review) on an infinite graph $G$  first requires 
to introduce the model on the Van Hove spheres $S_r$. 
For each site $i\in S_r$ $s_i$ is a q state function $s_i=1\dots q$ and 
the Hamiltonian is given by:
\begin{equation}
\label{defHr}
H_r= \sum_{(i,j)\in E_{r}} (1-\delta(s_i, s_j))+ h 
\sum_{i\in S_{r}} (1-\delta(s_i, 1)) 
\end{equation}
where the Kronecker delta-function is $\delta(s_i, s_j)=1$ if $s_i = s_j$ and 
$\delta(s_i, s_j)=0$ otherwise. Notice that the case $q=2$ is the Ising model.
In the canonical ensemble the partition function $Z_r$ is given by
the sum of the Boltzmann weight $\exp(-\beta H_r)$ 
over all possible configurations  $\{s_i \}$; $\beta$ represents the 
inverse temperature of the system. 
Thermodynamic properties of statistical models 
are described by extensive order parameters, in this
case the average magnetization:
\begin{eqnarray}
M_r(\beta,q,h) & = & 
|S_r|^{-1}\sum_{i\in S_r} \langle \delta(s_i,1)-q^{-1}\rangle_r \equiv \nonumber\\
& ~ & |S_r|^{-1} \sum_{i\in S_r} Z_r^{-1} \sum_{s_1 \dots s_{S_r}} 
(\delta(s_i,1)-q^{-1}) e^{-\beta H_r} ~.
\label{defMr}
\end{eqnarray}
where $\langle \cdot \rangle_r$ denotes the thermal average.
The Potts model on $G$ presents 
spontaneous magnetization if it 
exists $\beta_c$ such that for all $\beta > \beta_c$
\begin{equation}
\label{defM}
\lim_{h\to 0} M(\beta,q,h)\equiv 
\lim_{h\to 0} \lim_{r \to \infty} M_r(\beta,q,h) >0~.
\end{equation}
The existence of the thermodynamic limit $r\to\infty$ is always assumed in this
article.

Percolation (see e.g. \cite{F2,aharony} for a mathematical and a physical
treatment) is defined by introducing a probability $p$, $0\leq p\leq1$ and
each link $(i,j)\in E$ is declared to be open with probability $p$ and closed
with probability $1-p$ independently. The cluster $C_i$ containing the site $i$ 
is the set of all the sites connected with $i$ by a path of open links. 
The i-size of the cluster $C_i$ is the maximum chemical distance 
from $i$ of the sites of $C_i$. We call $P_i(l,p)$ the probability for
$i$ to belong to a cluster of i-size $\geq l$. A
graph $G$ is said to present (local) percolation if it exists a probability 
$p<1$ such that $\lim_{l\to\infty}P_i(l,p)>0$. On a graph with bounded
connectivity this property is independent of $i$ \cite{F2}.

A graph 
presents percolation on the average if it exists a probability $p$ such that
\begin{equation}
\label{defP}
\lim_{l\to \infty} \overline{P(l,p)} \equiv 
\lim_{l\to \infty} \lim_{r \to \infty} |S_r|^{-1} \sum_{i\in S_r} P_i(l,p) >0~.
\end{equation}
Percolation on the average is a new combinatorial problem and it gives rise to 
a graph classification which turns out to be fundamental for understanding the
thermodynamic behaviour of physical models on graph. In the main theorem of 
the paper, indeed, we prove that the q-states Potts model is spontaneously 
magnetized if 
and only if it is defined on a graph presenting percolation on the average,
obtaining a generalization to inhomogeneous structures of the classical result
for lattices by Aizenman et al. \cite{Aiz}.
The proof is mainly inspired by \cite{Hagg} and \cite{F1}. 

First, following
\cite{F1}, we introduce a representation of the q-states Potts model in terms 
of the q-random cluster model on the supplemented graph $G'$. Then, from an
important property of the random cluster stated in \cite{Hagg}, we get an 
inequality between the local magnetization and the percolation probability 
on the 
supplemented graph: this is an extension of \cite{Hagg} to the case $h\not=0$. 
In the second part of the proof we take the thermodynamic
average and we show that the physical requirements on the
graph allow to formulate the inequality for the average magnetization in terms 
of average percolation on $G$. The theorem directly follows from this 
inequality.

Let us first show that local and average percolation define different
classifications of inhomogeneous networks. In particular, the brush graph 
(Figure 1) presents local percolation but no percolation on the average. 
If we choose $p$ larger than the threshold of the two-dimensional 
percolation and we call $r$ the distance of $i$
from the plane of the brush, we have that 
$P_i(l,p)>P_{sq}(l-r,p) p^r$ ($P_{sq}(l,p)$ is the probability that a site of a
two-dimensional lattice belongs to a cluster of size $>l$) and then 
$\lim_{l\to\infty}P_i(l,p)>p^r \lim_{l\to\infty}P_{sq}(l-r,p)>0 $. 
On the other hand for
all the sites of the brush at a distance from the plane greater than $l$ we have
$P_i(l,p)=2p^l-p^{2l}$, $2p^l-p^{2l}$ is the probability for a 
site of a linear chain to belong to a cluster of i-size greater than $l$. 
We call $R_l$ the subset of the brush graph given by the sites whose 
distance from the plane is smaller than $l$. We have $||R_l||=0$ and 
\begin{equation}
\label{brush}
\lim_{l\to \infty} \overline{P(l,p)} \leq
\lim_{l\to \infty} ||R_l|| + (1-||R_l||) (2p^l-p^{2l}) =0~.
\end{equation}
Therefore, on the brush graph the q-states Potts model does not
have a single Gibbs measure \cite{Hagg}, however the system is not spontaneously magnetized
and the thermodynamic functions are analytical.

\section{Random cluster representation of the Potts model}

Let us define the supplemented q-random cluster model \cite{F1}. We call $G'$ 
the graph obtained by adding to $G$ a supplementary site $o$ connected by a new 
link to each site of $G$. The supplemented spheres are obtained by 
adding to the spheres $S_r$ the new vertex $o$ and the relevant edges. 
We will call $S'_r$ and $E_r'$ the set of sites and the set of links of the
supplemented sphere. The configurations $\xi$ of
the supplemented random cluster model on $S'_r$ are obtained by 
declaring each link of $E'_r$ open or closed. 
The supplemented q-random cluster model \cite{F1} is defined by choosing
each configuration according the probability distribution:
\begin{equation}
\label{defRC}
\mu(p,p_o,q,\xi)=p^{N_S(\xi)}\cdot (1-p)^{|E_r|-N_S(\xi)}\cdot p_o^{N_o(\xi)} 
\cdot (1-p_o)^{|S_r|-N_o(\xi)} q^{C(\xi)}
\end{equation}
where $p$, $p_o$ and $q$ are real parameters of the model ($q\geq 1$, $0\leq
p\leq 1$ and $0\leq p_o\leq 1$. $N_S(\xi)$ is the 
number of open links 
in $E_r$, $N_o(\xi)$ the number of open links connecting $o$ to $S_r$ and 
$C(\xi)$ the number of clusters in the configuration ($|S_r|$ and $|E_r|$
are respectively the cardinality of $S_r$ and $E_r$). For $q=1$ and $p_o=0$ 
the random cluster model is exactly the percolation model restricted to the
sphere $S_r$. 

As already pointed out in the original work by 
Fortuin and Kasteleyn \cite{F1}, q-states Potts model and q-random cluster model
are equivalent on any graph $G$. 
Indeed if we fix $p=1-e^{-\beta}$ and 
$p_o=1-e^{-\beta h}$, their partition functions coincide and 
\begin{equation}
\label{pottsrndclus}
\langle \delta(s_i,1)-q^{-1}\rangle_r=P_i^o(p,p_o,q)\equiv
\int F^o_i(\xi) \mu(p,p_o,q,\xi)
\end{equation}
where $F^o_i(\xi)=1$ if $o\in C_i$ and 
$F^o_i(\xi)=0$ otherwise. Then, $P_i^o(p,p_o,q)$ represents
the probability  for the site $i$ to belong to the same cluster of the 
supplemented site $o$.

The space of configurations of the random-cluster model is equipped by the 
partial order: $\xi \preceq \eta$ if the set of open links in $\xi$ is 
included or equal to the set of open links in $\mu$. 
We will say that the probability distribution $\mu(\xi)$ stochastically 
dominates the probability distribution $\mu'(\xi)$ ($\mu'\preceq_D \mu$) if:
\begin{equation}
\label{stcdom}
\int f d\mu' \leq \int f d\mu
\end{equation}
for all increasing function $f$.

In \cite{Hagg} it is proved that important inequalities for the probability
distributions follow from FKG \cite{FKG} 
theorem. These inequalities can be easily generalized to the case of 
supplemented graph, obtaining:
\begin{equation}
\mu(p',p'_o,1,\xi)\preceq_D \mu(p,p_o,q,\xi)\preceq_D \mu(p,p_o,1,\xi)
\label{ineq1}
\end{equation}
where $p'=p(q-p(q-1))^{-1}$ and $p_o'=p_o(q-p_o(q-1))^{-1}$.

Since $F^o_i(\xi)$ is an increasing function, from 
(\ref{pottsrndclus}) and (\ref{ineq1}) we have:
\begin{equation}
\label{ineq2}
P_i^o(p',p_o',1)\leq \langle \delta(s_i,1)-q^{-1}\rangle_r \leq P_i^o(p,p_o,1)
\end{equation}
(\ref{ineq2}) is an extension to the case $h\not=0$ of the inequality for local
magnetization proved in \cite{Hagg}.

\section{Spontaneous magnetization and percolation on the average}

Let us first take the average of 
(\ref{ineq2}) over all the sites of the sphere $S_r$. We obtain
\begin{equation}
\label{ineq3}
|S_r|^{-1} \sum_{i\in S_r} P_i^o(p',p_o',1)\leq M_r(\beta,q,h) \leq 
|S_r|^{-1} \sum_{i\in S_r}  P_i^o(p,p_o,1)
\end{equation}
Inequalities (\ref{ineq3}) provide 
upper and lower bounds for the magnetization of the Potts model in
terms of the percolation model defined on the supplemented Sphere $S'_r$. 
Let us reformulate these bounds in terms 
of the percolation on $G$.

First we prove a property of graphs with bounded connectivity and 
satisfying (\ref{isop}). In the probability measure defined on $S_r$ by 
(\ref{defRC}) with $q=1$ and $p_o=0$, we call $P_{i,r}(l,p)$ the
probability for the site $i$ to belong to a cluster of i-size greater than
$l$. Let us show that in the thermodynamic averages one can substitute 
$P_{i,r}(l,p)$  with the probability defined on the infinite graph $P_i(l,p)$.
We will call $S_{l,r}^+$ the subset of $S_r$ of all the sites $i$
such that the distance of $i$ from the border $\partial S_r$ is larger than $l$
and $S_{l,r}^-$ its complementary in $S_r$. For the sites $i\in S_{l,r}^+$ all 
the 
clusters $C_i$ of i-size smaller than $l$ are included in $S_r$, then the
probability of belonging  (or not belonging) to one of these clusters is equal 
in percolation on $G$ and in percolation on $S_r$.
Furthermore we have that:
\begin{equation}
||S_{l,r}^-||=\lim_{r\to \infty}|S_r|^{-1}|S_{l,r}^-|\leq 
\lim_{r \to \infty} |S_r|^{-1}|\partial S_r| z_{max}^{l+1}=0
\label{misS-}
\end{equation}
where we used (\ref{isop}) and the boundedness of the coordination number. 
Then $P_{i,r}(l,p)$ and $P_i(l,p)$ differ only on a set of zero measure and
in the thermodynamic averages we can exchange one for the other.

Let us now prove the main theorem of the paper.
From the boundedness of the correlation number and the independence 
of percolation probabilities, one gets the inequality:
\begin{equation}
\label{ineq4}
P_i^o(p,p_o,1)\leq \left(1-(1-p_o)^{z_{max}^l}\right)(1-P_{i,r}(l,p))+
P_{i,r}(l,p)~.
\end{equation}
Indeed $(z_{max}^l-1)/(z_{max}+1)$ is the maximum number of sites in a cluster 
of i-size $< l$ and then
$(1-(1-p_o)^{z_{max}^l})$ is an upper bound on the probability that a site of 
the cluster is connected to $o$.
From (\ref{ineq3}) and (\ref{ineq4}) taking the thermodynamic limit 
$r\to \infty$  we have:
\begin{equation}
M(\beta,q,h)  \leq  \lim_{r \to \infty}
{1 \over |S_r|}
\sum_{i\in S_r}\left( \left(1-(1-p_o)^{z_{max}^l}\right)
(1-P_i(l,p))+P_i(l,p) \right) ~.
\label{ineq5}
\end{equation}
Letting $h \to 0$, since also $p_o \to 0$, we get
\begin{equation}
\lim_{h\to 0}M(\beta,q,h)
\leq \lim_{h \to 0}\left(1-(1-p_o)^{z_{max}^l}\right) +\overline{P(l,p)}
=\overline{P(l,p)}~.
\label{ineq6}
\end{equation}
Now we have to consider the first inequality in (\ref{ineq3}). From the
probability independence in percolation, we get 
\begin{equation}
\label{ineq7}
P_i^o(p',p'_o,1)\geq \left (1-(1-p'_o)^{l+1}\right) P_{i,r}(l,p')~.
\end{equation}
Here $l+1$ is the minimum number of sites in a cluster of i-size $\geq l$,
and then 
$(1-(1-p'_o)^{l+1})$ is a lower bound on the probability that a site of the 
cluster is connected to $o$. Taking the thermodynamic average we have
\begin{equation}
M(\beta,q,h)  \geq  \left (1-(1-p'_o)^{l+1}\right) \overline{P(l,p')}~.
\label{ineq8}
\end{equation}
Letting first $l \to \infty$ and then $h\to 0$ we have
\begin{equation}
\lim_{h\to 0} M(\beta,q,h)  \geq  \lim_{l \to \infty} \overline{P(l,p')}
\label{ineq9}
\end{equation}
Finally, from (\ref{ineq6}) and (\ref{ineq9}) one has
\begin{equation}
\lim_{l \to \infty} \overline{P(l,p')} \leq
\lim_{h\to 0} M(\beta,q,h)  \leq  \lim_{l \to \infty} \overline{P(l,p)}
\label{ineq10}
\end{equation}
If the graph presents percolation on the average, it exists a probability $p'$ 
such
that $\lim_{l \to \infty} \overline{P(l,p')}>0$ and then the first inequality 
in (\ref{ineq10}) implies that the model for $\beta=\log(1+p'q(1-p')^{-1})$ is
magnetized. On the other hand, if $\lim_{l \to \infty} \overline{P(l,p)}=0$ for
all values of $p$, the second inequality in (\ref{ineq10}) implies that the
magnetization of the system is zero for all temperatures and this
concludes the proof.

The result can be easily generalized to Potts models with disordered
ferromagnetic couplings described by
the Hamiltonian $H_r= \sum_{(i,j)\in E_{r}} J_{ij}(1-\delta(s_i, s_j))+ h 
\sum_{i\in S_{r}} (1-\delta(s_i, 1)) $ with $0<\epsilon<J_{ij}<K$ for all
$(i,j)\in G$.

On the other hand an important open question is the possibility to extend the 
proof to different ferromagnetic spin models with discrete 
symmetry, such as the clock model.
As mentioned in the introduction,
another basic problem is a better understanding of the relation  between
percolation on the average, transience on the average and graph topology.

\ack

The author is grateful to R. Burioni and D. Cassi for useful 
comments and discussions.

\end{document}